\def\alr{A_{LR}}
\def\Mol{M{\o}ller\ }
\def\ssw{s_W^2}
\def\sswq{s_W^4}

\def\sll{\sigma_{{\rm LL}}}
\def\slr{\sigma_{{\rm LR}}}
\def\srl{\sigma_{{\rm RL}}}
\def\srr{\sigma_{{\rm RR}}}

\def\dsll{{\rm d}\sigma_{{\rm LL}}}
\def\dslr{{\rm d}\sigma_{{\rm LR}}}
\def\dsrl{{\rm d}\sigma_{{\rm RL}}}
\def\dsrr{{\rm d}\sigma_{{\rm RR}}}

\def\nll{N_{{\rm LL}}}
\def\nlr{N_{{\rm LR}}}
\def\nrl{N_{{\rm RL}}}
\def\nrr{N_{{\rm RR}}}

\def\peff{P_{{\rm eff}}}

\def\msbar{{\overline{\rm MS}}}
\def\alr{A_{\rm LR}}

\def\cw{\cos^2\theta_W}
\def\sw{\sin^2\theta_W}
\def\swq{\sin^4\theta_W}

\def\ea{\end{eqnarray}}
\def\ba{\begin{eqnarray}}
\def\ee{\end{eqnarray}}
\def\be{\begin{eqnarray}}
\newcommand{\gsim}{\lower.7ex\hbox{$
\;\stackrel{\textstyle>}{\sim}\;$}}
\newcommand{\lsim}{\lower.7ex\hbox{$
\;\stackrel{\textstyle<}{\sim}\;$}}

\documentstyle[ee,twoside,psfig]{article}
\begin{document}
\normalsize\textlineskip

\begin{flushright}
BNL-HET-00/2 \\
hep-ph/0003049 \\ March 2000
\end{flushright}

\title{POLARIZED M{\O}LLER SCATTERING ASYMMETRIES\footnote{Talk given
at $e^-e^-99$: 3rd International Workshop on Electron-Electron
Interactions at TeV Energies, Santa Cruz, December 1999.}}
\author{ANDRZEJ CZARNECKI and WILLIAM J. MARCIANO} 
\address{
Brookhaven National Laboratory\\  Upton, New York 11973}

\maketitle\abstracts{
The utility of polarized electron beams for precision electroweak
studies is described. Parity violating M{\o}ller scattering
asymmetries in $e^-e^- \to e^-e^-$ are discussed.  Effects of
electroweak radiative corrections and the running $\sin^2\theta_W(Q^2)$
are reviewed.  The sensitivity of E158 (a fixed target $e^-e^-$
experiment at SLAC) and future $e^-e^-$ 
collider studies to ``new physics'' is briefly outlined.
}

\vspace*{5mm}

\section{Polarization and Precision Measurements}

Polarized beams provide powerful tools for testing the Standard Model
and probing ``new physics'' effects.  They can be used to enhance
signals, suppress backgrounds, study particle properties, and carry
out precision measurements.  A beautiful illustration of the last
possibility is provided by the SLD measurement of $A_{LR}$ at the $Z$ pole
\ba
\alr \equiv {\sigma\left( e^+e^-_L \to {\rm hadrons}\right)
-\sigma\left( e^+e^-_R \to {\rm hadrons}\right)
\over 
\sigma\left( e^+e^-_L \to {\rm hadrons}\right)
+\sigma\left( e^+e^-_R \to {\rm hadrons}\right)
}.
\label{eq1}
\ea
That quantity is very sensitive to $\sw$
\ba
\alr = {2(1-4\sw)\over 1+(1-4\sw)^2}  \qquad \mbox{(Tree level)}.
\label{eq2}
\ea
In fact, for $\sw \simeq 0.23$, one finds $\Delta\sw/\sw \simeq
-{1\over 10}\Delta \alr /\alr$.  Hence, a $\pm 1\%$ measurement of
$\alr$ determines $\sw$ at the $\pm 0.1\%$ level.

Based on about 500 thousand $Z$ decays and employing a polarized
$e^-$ beam with polarization reaching $P_{e^-}\simeq 77\%$, the SLD
collaboration has reported\cite{Swartz:1999xv} the single best
measurement of the weak 
mixing angle  (defined here by modified minimal subtraction)
\ba 
\sw(m_Z)_{\overline{MS}} = 0.23073\pm 0.00028,
\label{eq3}
\ea
which weighs heavily in the (leptonic) $Z$ pole average (from SLD and LEP)
\ba
\sw(m_Z)_{\overline{MS}} = 0.23091\pm 0.00021.
\label{eq4}
\ea

Taken on their own, the quantities in eqs.~(\ref{eq3}) and (\ref{eq4})
are merely precise numbers. They become interesting when interpreted
in the context of a complete (renormalizable) theory such as the
$SU(3)_C\times SU(2)_L \times U(1)_Y$ Standard Model or its various
extensions. Then, symmetries provide natural relationships  among
couplings and masses which can be tested by comparing different
precision measurements.  For example, the fine structure constant,
Fermi constant, and $Z$ mass
\ba
\alpha^{-1} &=& 137.03599959(40) 
\nonumber \\
G_\mu &=& 1.16637(1) \times 10^{-5} \; \mbox{GeV}^{-2}
\nonumber \\
m_Z &=& 91.1871(21) \; \mbox{GeV}
\label{eq5}
\ea
can be compared with the weak mixing angle via
\ba
\sin^22\theta_W(m_Z)_{\overline{MS}} = {4\pi\alpha \over \sqrt{2}G_\mu
m_Z^2
\left[ 1-\Delta\hat r (m_t, m_h)\right]}
\label{eq6}
\ea
where $\Delta\hat r$ represents finite, calculable quantum loop
effects which depend on the top quark and Higgs scalar masses.  Taking
$m_t=174.3 \pm 5.1$ GeV and $m_h \simeq 100$ GeV leads to $\Delta\hat
r = 0.05940\pm 0.0005\pm 0.0002$, where the errors correspond
to $\Delta m_t$ and hadronic loop uncertainties.

Leaving $m_h$ arbitrary, eq.~(\ref{eq6}) leads to the
prediction\cite{Degrassi:1997iy} 
\ba
\sw (m_Z)_{\overline{MS}} = (0.23112\pm 0.00016 \pm 0.00006)
\left( 1+0.00226\, \ln {m_h\over 100 \;\mbox{GeV}}\right).
\label{eq7}
\ea
Comparing that prediction with the world average in eq.~(\ref{eq4})
suggests a relatively light Higgs,
\ba
m_h\simeq 65
{\scriptsize
\begin{array}{@{\hspace*{-0.1mm}}r@{\hspace*{0.3mm}}r@{\hspace*{0.3mm}}r}
+35 & +28 & +9\\
-20 & -21 & -8
\end{array}
}
\; \mbox{GeV},
\label{eq8}
\ea
which is centered somewhat below the LEP II direct search
bound\cite{wu99} 
\ba
m_h > 106 \; {\rm GeV} \qquad \mbox{(95\% C.L.)}.
\label{eq9}
\ea
In fact, the SLD value in eq.~(\ref{eq3}) favors an even smaller
$m_h$.  If the Higgs mass turns out to be well outside the range in
eq.~(\ref{eq8}), then one must append ``new physics'' to the Standard
Model either through loop effects or small tree level contributions.  

It would be nice to push the current $\pm 0.1\%$ test in
eq.~(\ref{eq6}) as far as possible.  Indeed, $\alpha$, $G_\mu$, and
$m_Z$ are all already known to much better than $\pm 0.01\%$ (and will be
or can be further improved).  Can one reduce the uncertainty in
$\sw(m_Z)_\msbar$ from its current $\pm 0.1\%$ to $\pm 0.01\%$?  If
so, it would provide a sensitivity to $m_h$ at the incredible $\pm
5\%$ level (assuming $m_t$ and hadronic loop uncertainties are also
improved).

The only known way to improve $\sw(m_Z)_\msbar$ is to carry out a
clean high statistics study of asymmetries such as $\alr$.  In that
regard, the NLC (Next Linear Collider) will be capable at an early
stage of sitting at
the $Z$ resonance and collecting $10^8 - 10^9$ $Z$ decays in a
relatively short time.  With such statistics, $\sw (m_Z)_\msbar$ can,
in principle,
be obtained via $\alr$ to better than $\pm 0.01\%$. Systematics then
become the issue.  The dominant systematic uncertainty at the SLD was
a $\pm 0.5\%$ polarization error which contributes to $\Delta \sw$ at
the $\pm 0.0001$ level.  One would need to reduce the polarization
uncertainty to $\pm 0.1\%$ to reach $\pm 0.01\%$ in $\sw
(m_Z)_\msbar$.  Such a reduction would be possible if both the $e^+$
and $e^-$ beams were polarized.  Then, the effective polarization
(they add like relativistic velocities)
\ba
P_{\rm eff} = {P_{e^-}-P_{e^+}\over 1 -P_{e^-}P_{e^+}}
\label{eq10}
\ea
enters
\ba
{N_{LR}-N_{RL} \over N_{LR}+N_{RL}}
= P_{\rm eff} \alr,
\label{eq11}
\ea
where $N_{LR}$ denotes the number of $e^-_Le^+_R$ induced hadronic $Z$
decays.  
For $|P_{e^-}|=0.9000\pm 0.0045$ and $|P_{e^+}|=0.6500\pm 0.0065$
(i.e. $\pm 1\%$ $e^+$ polarization), one finds $P_{\rm eff} =
0.9779\pm 0.0012$ as required for a $\pm 0.01\%$ determination of
$\sw(m_Z)_\msbar$. 

Improving the direct 
measurement of $\sw(m_Z)_\msbar$ can have other
applications.  The $Z$ pole determination is relatively pure and free
of ``new physics.''  Below, we demonstrate its utility for comparison
with polarized M{\o}ller scattering asymmetries which could exhibit
effects from ``new physics'' beyond the Standard Model.

\section{Polarized M{\o}ller Scattering -- Fixed Target}
\begin{figure}[htb]
\noindent
\begin{minipage}{16.cm}
\hspace*{-5mm}
$
\mbox{
\hspace*{22mm}
\begin{tabular}{ccc}
\psfig{figure=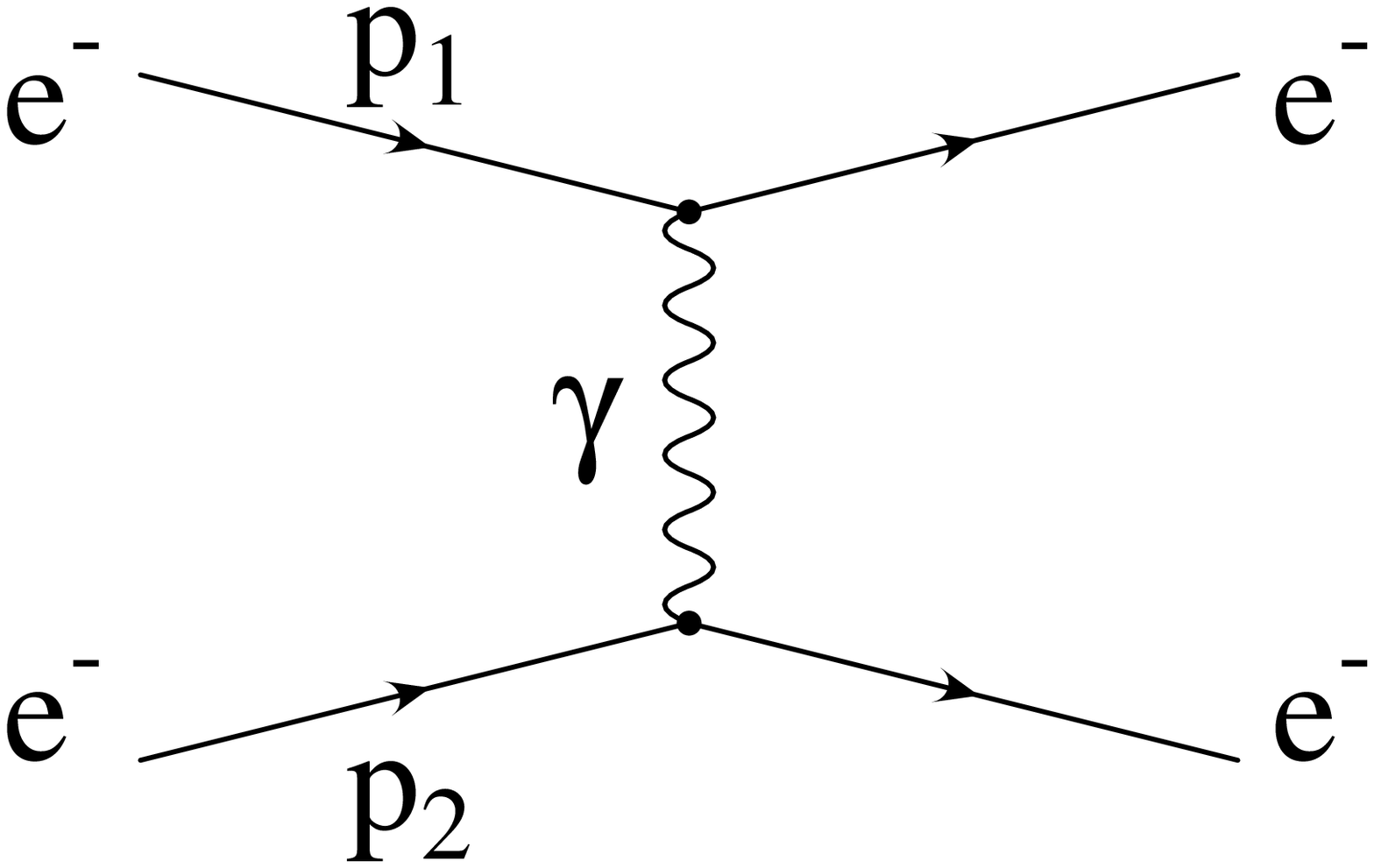,width=3.8cm,bbllx=210pt,bblly=410pt,%
bburx=630pt,bbury=550pt} 
&+&\hspace*{18mm}
\psfig{figure=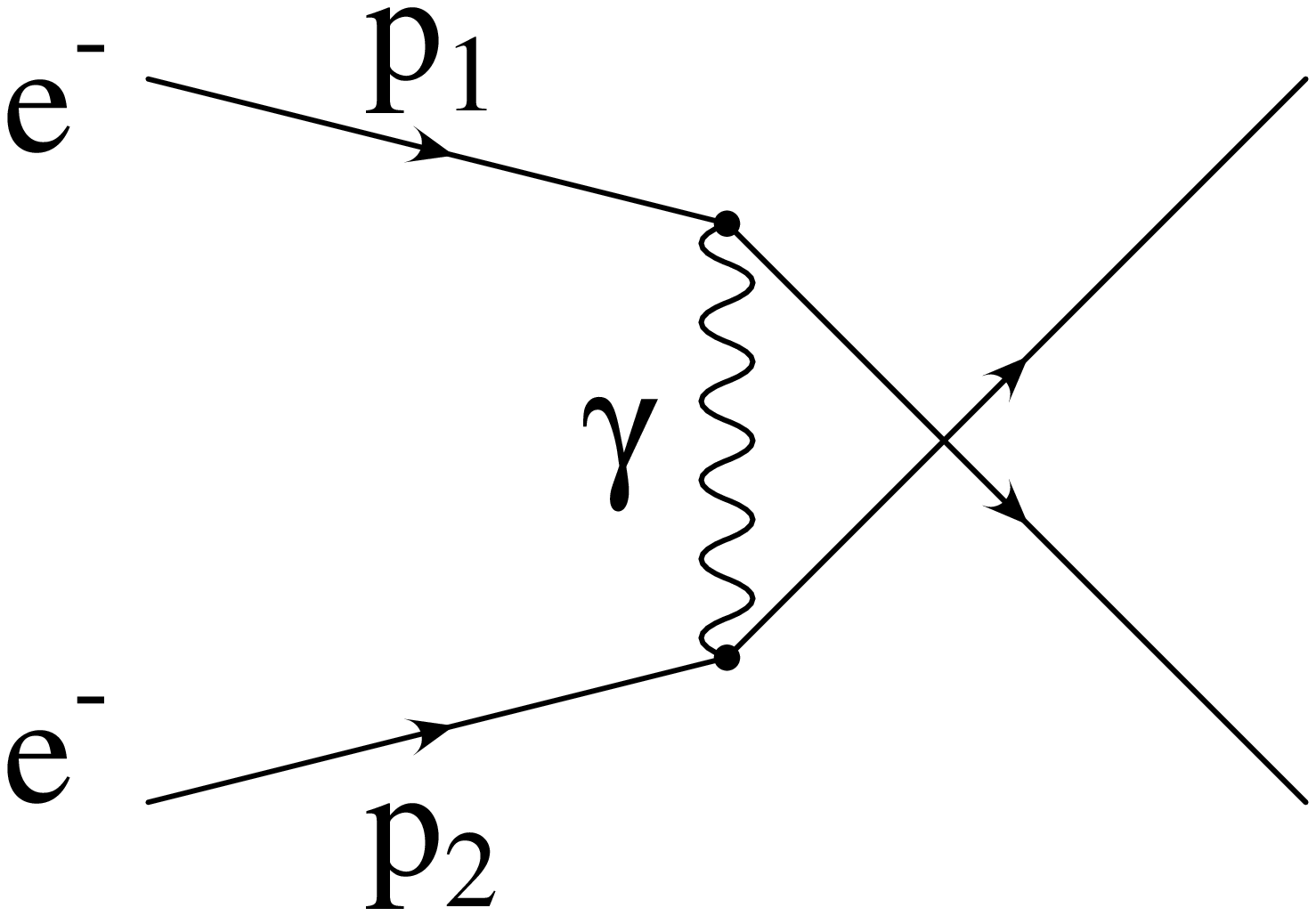,width=3.8cm,bbllx=210pt,bblly=410pt,%
bburx=630pt,bbury=550pt} \\[2cm]
\psfig{figure=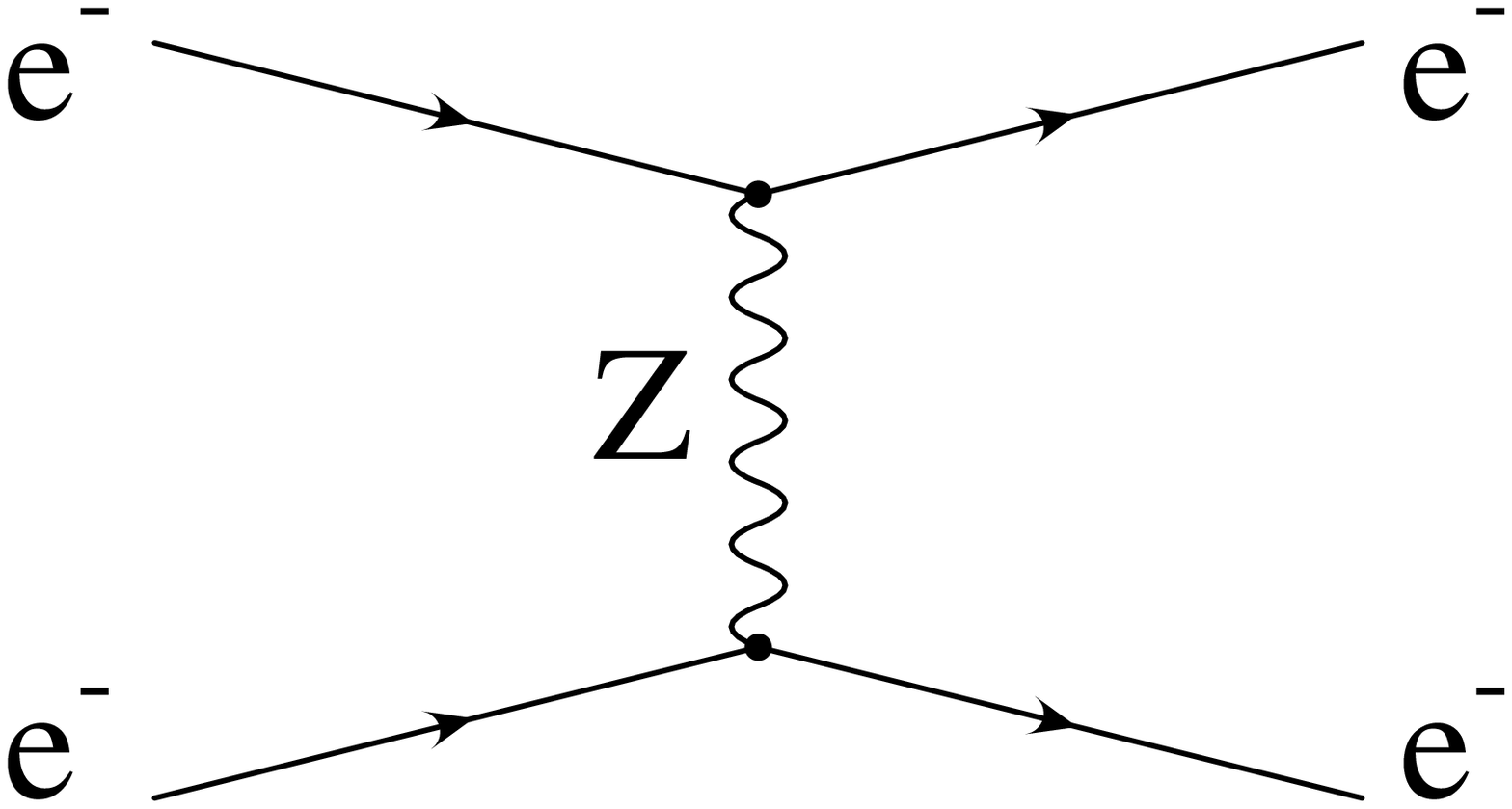,width=3.8cm,bbllx=210pt,bblly=410pt,%
bburx=630pt,bbury=550pt}
&+&\hspace*{18mm}
\psfig{figure=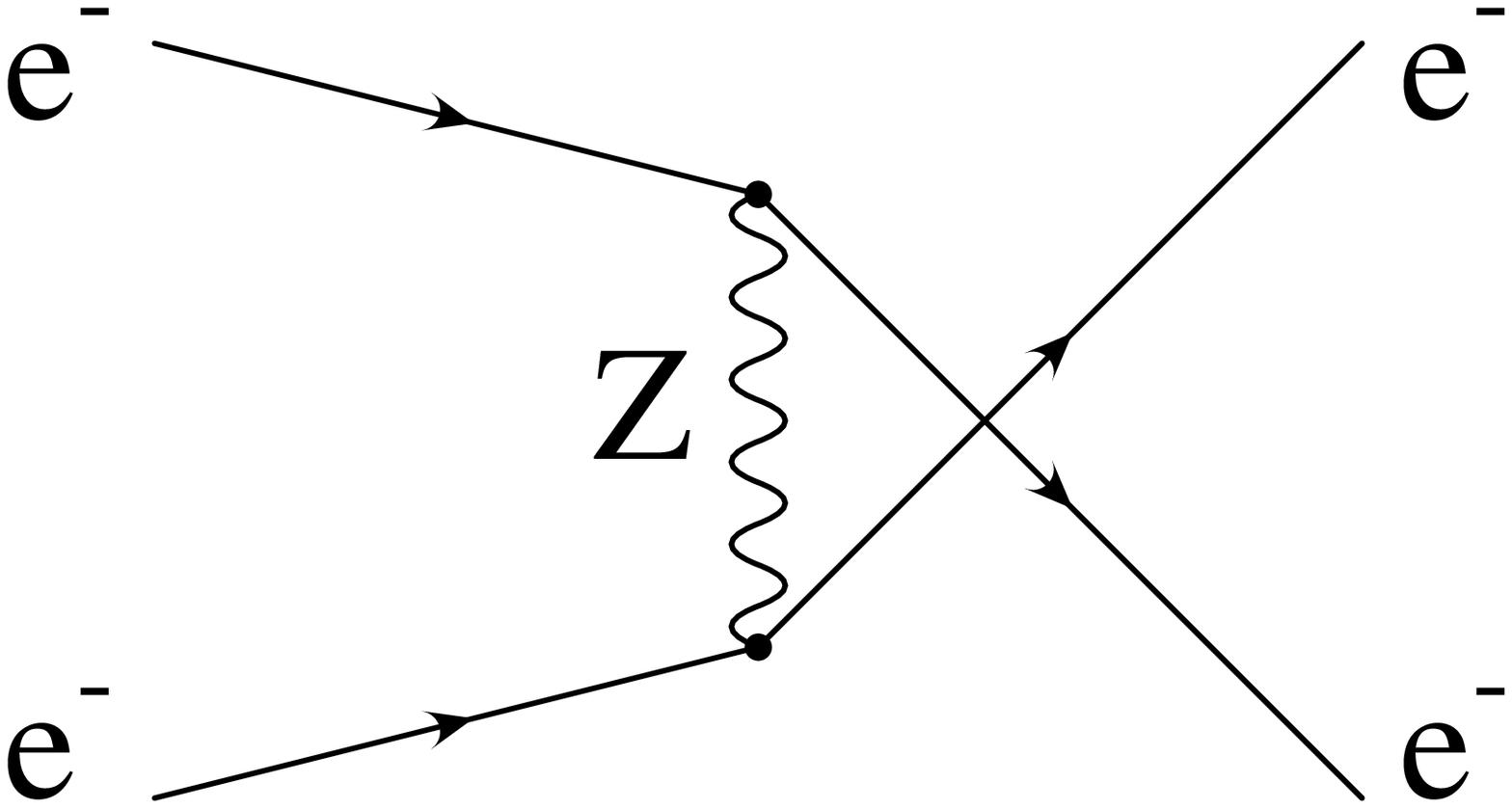,width=3.8cm,bbllx=210pt,bblly=410pt,%
bburx=630pt,bbury=550pt} 
\end{tabular}
}
$
\end{minipage}

\vspace*{15mm}
\noindent 
\fcaption{Neutral current direct and crossed $e^-e^-$ scattering 
amplitudes leading to the asymmetry $A_{LR}$ at tree level.}
\label{fig:diagrams}
\end{figure}

M{\o}ller scattering $e^-e^-\to e^-e^-$ has been a well studied, 
classic low energy reaction.\cite{moeller32}
Employing polarized electrons, one can, in principle,
measure parity violating weak
interaction 
asymmetries.\cite{tree}
At tree level, the $\alr$ in M{\o}ller scattering comes from an
interference among the diagrams in Fig.~{\ref{fig:diagrams}}.  
For a single polarized $e^-$, the asymmetry corresponds to 
\ba
\alr^{(1)} \equiv 
{\dsll+\dslr-\dsrl-\dsrr \over \dsll+\dslr+\dsrl+\dsrr}
\label{eq12}
\ea
while in the case of both $e^-$ polarized, a second asymmetry becomes
possible\cite{Cuypers:1996it}
\ba
\alr^{(2)} \equiv 
{\dsll-\dsrr \over \dsll+\dsrr}.
\label{eq13}
\ea
The subscripts denote the initial $e^-e^-$ states' polarizations.
As we subsequently show, both asymmetries
would be measurable at a high energy $e^-e^-$ collider where
polarizations of $0.90$ for each beam are likely.   Since
d$\sigma_{LR} = {\rm d}\sigma_{RL}$ by rotational invariance, they
differ only in their denominators.

Let us begin by considering a fixed target scenario in which a 50 GeV
polarized electron beam scatters off a fixed target of electrons.
That case will be addressed in the near future by SLAC experiment
E158.\cite{Kumar:1995ym}  

In the center-of-mass frame, the differential cross section is
characterized by the scattering angle $\theta$ with respect to the
beam axis or
\ba
y={1-\cos\theta\over 2}, \qquad 0\le \theta \le \pi.
\ea
The variable $y$ relates the momentum transfer $Q^2=-q^2$ and
center-of-mass energy $\sqrt{s}$ via
\ba
Q^2 = ys, \qquad 0\le y \le 1.
\ea
Since the cross section grows as $1/y^2(1-y)^2s$, very high statistics
are possible at small angle and/or small $s$.  However, the asymmetry
grows with $s$.  All things considered, it is generally better to
measure $\alr$ at high $s$, but lower energy fixed target facilities
can compensate by having very large effective luminosities.  For
example, 
E158 at SLAC will have $s\simeq 0.05\, {\rm GeV}^2$ and aims to
measure (with high precision) a very small asymmetry
$\alr \sim 1.5 \times 10^{-7}$.  That is
only possible because their luminosity will be ${\cal L}\simeq 4\times
10^{38} {\rm cm}^{-2}/s$.  

At small $Q^2 = ys \ll m_Z^2$, the left-right polarization asymmetry
in M{\o}ller scattering is given by (at tree level)\cite{tree}
\ba
\alr^{(1)}(e^-e^-\to e^-e^-) = {G_\mu s\over \sqrt{2}\pi\alpha}
{y(1-y)\over 1+y^4 +(1-y)^4} (1-4\sw),
\label{eq16}
\ea
or for comparison with the $Z$ pole asymmetry
\ba
\lefteqn{\alr^{(1)}(e^-e^-\to e^-e^-)}\nonumber \\
&& = {12\over \alpha}
{y(1-y)\over 1+y^4 +(1-y)^4} 
{s\Gamma(Z\to e^+e^-)\over m_Z^3}
\alr ( e^+e^- \to Z \to {\rm hadrons}).
\label{eq17}
\ea
To be at all competitive with the $\pm 0.00028$ uncertainty in $\sw$
found by SLD, very high statistics are required or equivalently, a
very good determination of $\alr$,
\ba
{\delta \sw\over \sw} \simeq -{1-4\sw \over 4\sw} {\delta \alr \over
\alr}.
\label{eq18}
\ea
Again, one sees the enhanced sensitivity to small changes in $\sw$.  
E158 aims for a $\pm 0.0007$ to $\pm 0.0004$ measurement of $\sw$
which will make it the best low energy determination of that
quantity.  As we subsequently illustrate, it will be sensitive
to the running of the weak mixing angle as well as ``new physics''
effects.

\section{Polarized M{\o}ller Scattering at Collider Energies}
\Mol scattering, $e^-e^-\to e^-e^-$, at the NLC can also be used for
precision tests of the Standard Model as well as direct and indirect 
searches for ``new
physics.''\cite{Heusch,Heusch97}  Indeed, in some cases it can provide
a more powerful probe 
than $e^+e^-$. One can assume with some confidence that both $e^-$
beams will be polarized with $|P_1|=|P_2|=0.9$ and about $\pm 0.5$\%
uncertainty each. The effective polarization will therefore be
(with like sign $P_1$ and $P_2$)
\ba
\peff = {P_1+P_2\over 1+P_1P_2} = 0.9945 \pm 0.0004.
\ea
We see that $\peff$ will be very large and has essentially
negligible uncertainty compared to  $P_1$ and $P_2$.  

The differential cross section in high energy collider 
\Mol scattering is also
characterized by a single parameter, the scattering angle $\theta$
with respect to the beam axis or
\ba 
y={1-\cos\theta\over 2}, \qquad 0\le \theta \le \pi.
\ea
The cross section grows as $1/y^2$ for small angle scattering. Hence,
very 
high statistics are possible in the small angle region. Good angular
coverage is therefore important for precision measurements.
As before, 
the variable $y$ relates $s$ and the momentum transfer $Q^2 = -q^2$
via 
$Q^2=ys, 0\le y\le 1$.
Note, that $y$ and $1-y$ correspond to indistinguishable events. Very
forward (small angle) 
$e^-e^-$ events will therefore be composed of high and low
$Q^2$ contributions.

As previously noted, one can consider two distinct but similar parity
violating \Mol 
asymmetries: the single spin asymmetry $\alr^{(1)}$ defined in
eq.~(\ref{eq12}) and double  spin asymmetry $\alr^{(2)}$ in
eq.~(\ref{eq13}).

Experimentally, one can and probably will flip the individual
polarizations (pulse by pulse) and measure $\nll$, $\nlr$, $\nrl$, and
$\nrr$ (the number of events in each mode) for fixed luminosity and
polarization.  From those measurements, the polarizations and
$\alr^{(2)}(y)$ can be simultaneously determined
using\cite{Cuypers:1996it,Czarnecki:1998xc} 
\be
{\nll+\nlr-\nrl-\nrr \over\nll+\nlr+\nrl+\nrr}&=&\phantom{-}P_1\alr^{(1)}(y),
\label{eq:18}
\\
{\nrr+\nlr-\nrl-\nll \over\nrr+\nlr+\nrl+\nll}&=&-P_2\alr^{(1)}(y),
\label{eq:19}
\\
{\nll-\nrr\over \nll+\nrr} &=& \peff \alr^{(2)}(y)
\left(
{1\over 1+ {1-P_1P_2\over 1+P_1P_2}{\slr+\srl\over \sll+\srr}}
\right),
\label{eq:20}
\\
\peff&=& {P_1+P_2\over 1+P_1P_2}.
\nonumber
\ee
For $P_1=P_2=0.9$, the correction term in parentheses of
Eq.~(\ref{eq:20}) is small but must be accounted for.  Using
Eq.~(\ref{eq:20}), $\alr^{(2)}$ (which depends on $\sw$) can be
extracted from data and compared with the Standard Model prediction. A
deviation from expectations would signal ``new physics.''

In general the d$\sigma_{ij}$ for \Mol scattering are somewhat lengthy
expressions\cite{Czarnecki:1998xc}  
with contributions from direct and crossed $\gamma $ and
$Z$ exchange amplitudes (see Fig.~1).
To simplify our discussion, we consider for illustration the
case $ys$ and $(1-y)s\gg m_Z^2$; so, terms of relative order $m_Z^2/ys
$ and  $m_Z^2/(1-y)s$ can be neglected. In that limit, one finds at
tree level\cite{Czarnecki:1998xc} 
\be
{\dsll\over {\rm d}y} &=& \sigma_0 {1\over y^2(1-y)^2} {1\over
16\swq},
\nonumber \\
{\dsrr \over {\rm d}y} &=&\sigma_0 {1\over y^2(1-y)^2},
\nonumber \\
{\dslr\over {\rm d}y} &=& {\dsrl\over {\rm d}y}= \sigma_0 {y^4+(1-y)^4
\over y^2(1-y)^2} 
{1\over 4},
\label{eq:21}
\ee
and the asymmetries become
\be 
\alr^{(1)}(y) &=& {(1-4\ssw)(1+4\ssw)
\over 
1+16\sswq + 8\left[y^4+(1-y)^4\right]\sswq},
\label{eq:22} \\
\alr^{(2)}(y) &=& {(1-4\ssw)(1+4\ssw)
\over 
1+16\sswq}.
\label{eq:23} 
\ee
Expanding about $\sw=1/4$, Eq.~(\ref{eq:23}) becomes
\be
\alr^{(2)}(y) = (1-4\sw)+{\cal O}[(1-4\sw)^2].
\ee
For arbitrary $s$, the asymmetries are maximal at $y=1/2$.  There 
we find, up to terms of ${\cal O}[(1-4\sw)^2]$,
\be
\alr^{(1)}(y=1/2) &\approx & (1-4\sw)
{\frac{16\,x\,\left( 3 + 2\,x \right) }{3\,\left( 27 + 34\,x +
11\,{x^2} \right) }}, 
\nonumber \\[2mm]
\alr^{(2)}(y=1/2) &\approx & (1-4\sw){\frac{2\,x}{3 + 2\,x}},
\qquad x\equiv {s\over m_Z^2}.
\ee

Because of the $(1-4\sw)$ dependence of
$\alr(e^-e^-)$, even with relatively modest angular coverage limited
to $0.1\le y\le 0.9$, \Mol scattering can be used to measure $\sw$
rather precisely, to about $\pm 0.0003$ at $\sqrt{s}\approx 1$
TeV. Although not likely to compete with future potential very high
statistics $Z$ pole measurements, it will be competitive with present
day measurements.  In addition, \Mol scattering can be used as a
powerful probe for ``new physics'' effects.  Indeed, for electron
composite effects parametrized by the four fermion
interaction\cite{Barklow:1996ut} 
${2\pi\over \Lambda^2}  \overline{e}_L \gamma_\mu e_L
 \overline{e}_L \gamma^\mu e_L$
one finds $\Delta\alr \approx
sy(1-y)c_W^2/\alpha\Lambda^2$ for $e^-e^-$ \Mol scattering.  It can,
therefore, be more sensitive than $e^+e^-\to e^+e^-$  
(about 50\% better) and could probe $\Lambda
\sim 150$~TeV. 

If one is interested in an even more precise determination of $\sw$ via \Mol
scattering, extremely forward events must be detected. For example,
assuming detector acceptance down to about $5^\circ$ ($y=0.0019$),
Cuypers and Gambino\cite{Cuypers:1996it} have shown that $\Delta\sw
\approx \pm 0.0001$ may be possible at a $\sqrt{s}=2$ TeV $e^-e^-$
collider with $P_1=P_2=90\%$.

\begin{figure}
\noindent
\begin{minipage}{16.cm}
\hspace*{14mm}
$
\mbox{
\begin{tabular}{cc}
\psfig{figure=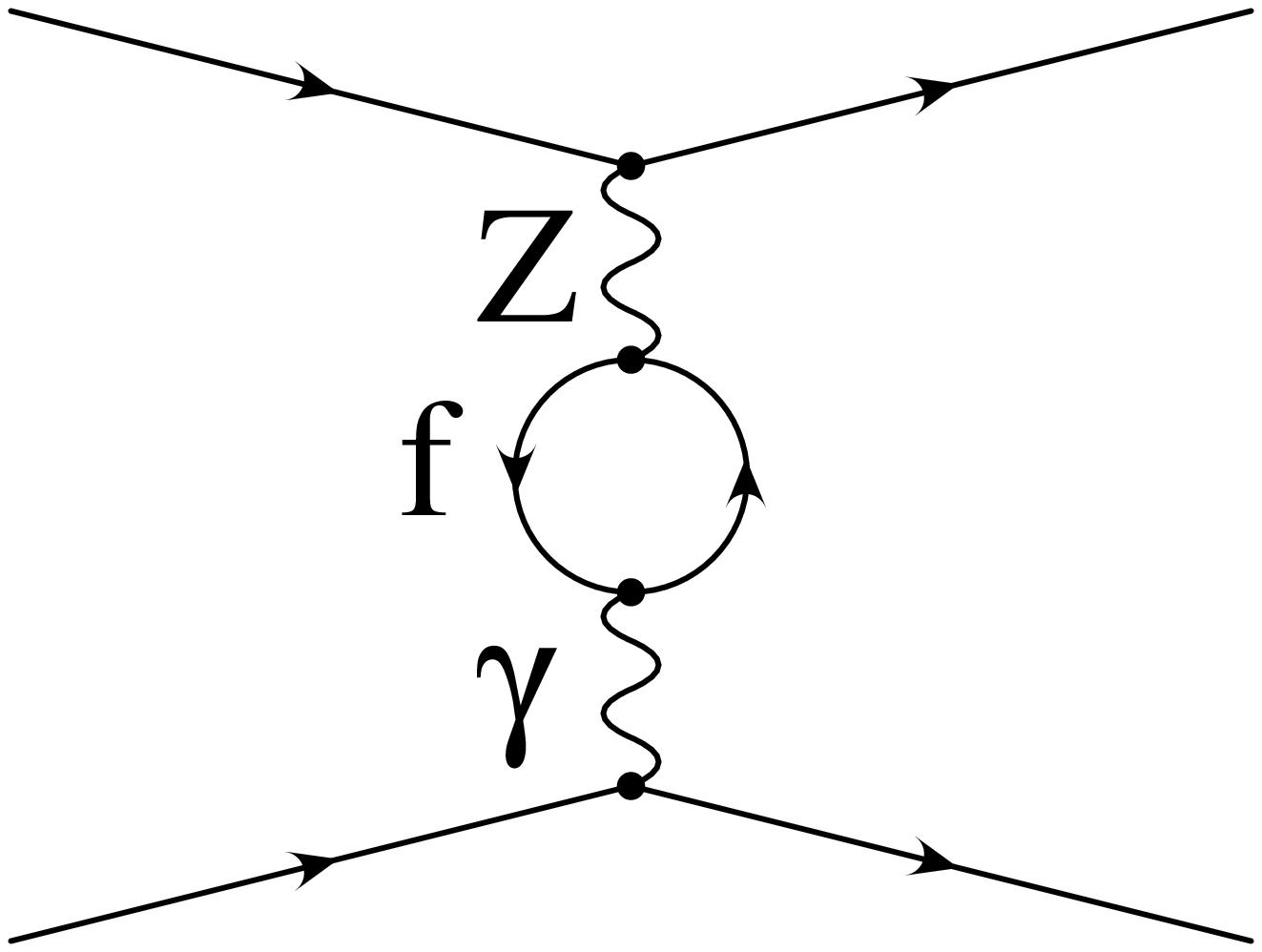,width=4.2cm,bbllx=210pt,bblly=410pt,%
bburx=630pt,bbury=550pt} &\hspace*{15mm}
\psfig{figure=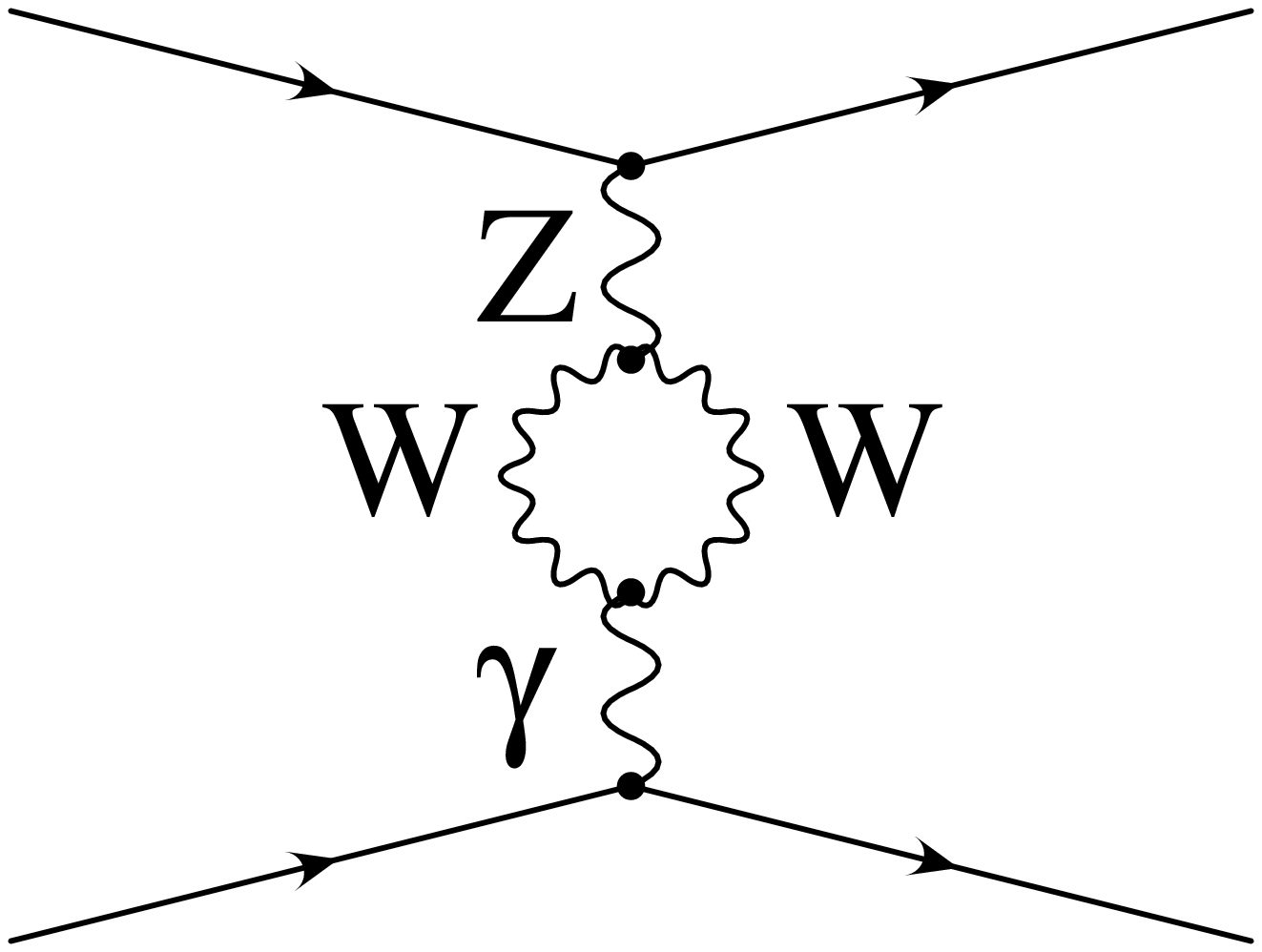,width=4.2cm,bbllx=210pt,bblly=410pt,%
bburx=630pt,bbury=550pt}
\\[20mm]
\psfig{figure=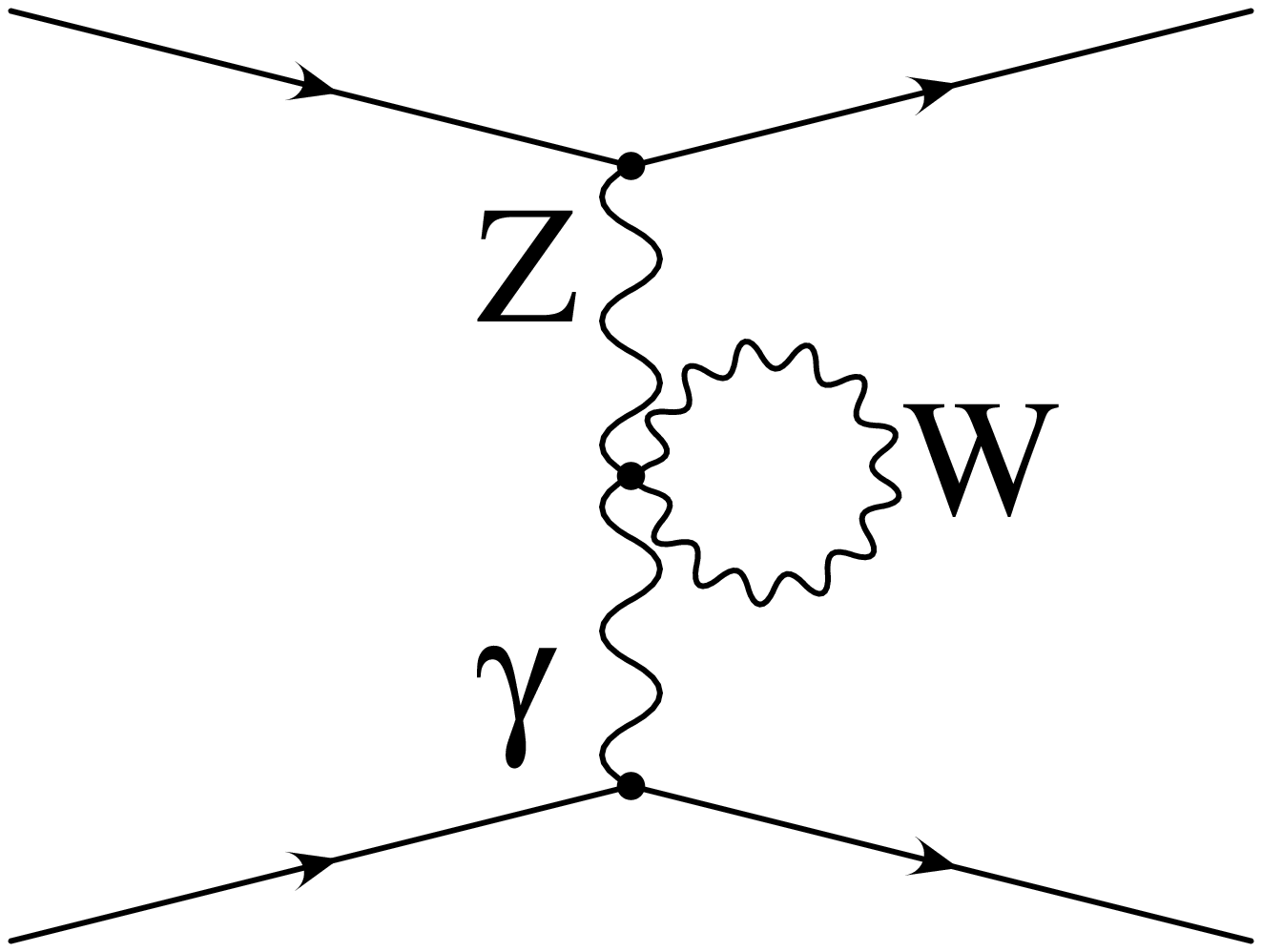,width=4.2cm,bbllx=210pt,bblly=410pt,%
bburx=630pt,bbury=550pt} &\hspace*{15mm}
\psfig{figure=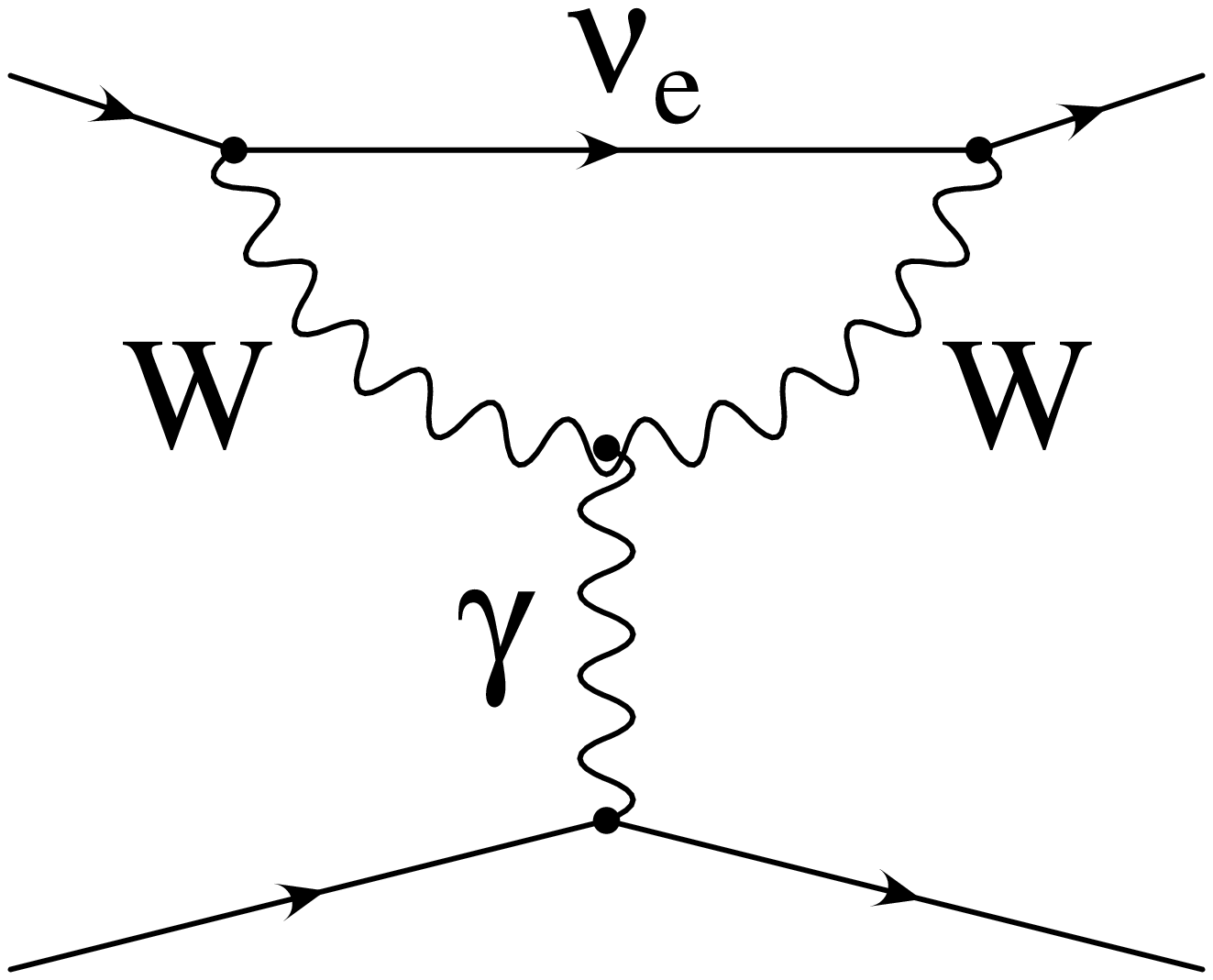,width=4.2cm,bbllx=210pt,bblly=410pt,%
bburx=630pt,bbury=550pt}\\[12mm]
\end{tabular}
}
$
\end{minipage}

\vspace*{4mm}
\noindent 
\fcaption{  $\gamma-Z$ mixing diagrams and  $W$-loop contribution to
the anapole moment.}
\label{fig:ew}
\end{figure}

\section{Radiative Corrections and \boldmath $\sw(Q^2)$}
The tree level $\alr$ for both E158 and future $e^-e^-$ collider
studies are 
proportional to $1-4\sw$ and hence suppressed because $\sw \simeq
0.23$.  Since some electroweak radiative corrections are not
suppressed by $1-4\sw$, they can be potentially very large.  A
complete calculation has been carried out\cite{Czarnecki:1996fw} for
small $s$ as appropriate to E158. 
There it was shown that such effects reduce $\alr$ by 40\% and must be
included in any detailed study.  Here, we comment on the primary
sources of those large corrections and show how much of the effect can
be incorporated into a running $\sw(Q^2)$.  We also discuss how those
large effects carry over to collider energies.  For a complete study
of radiative corrections to M{\o}ller scattering at high energies, see
ref.~\citenum{Denner:1998um}, \citenum{Jadach:1996im}, 
\citenum{Shumeiko:1999zd}.

The largest radiative corrections to $\alr$ at low energies come from
three sources:
\begin{enumerate}
\item $WW$ box diagrams,
\item Photonic vertex and box diagrams,
\item $\gamma Z$ mixing and the anapole moment.
\end{enumerate}
The first two are of order $+4\%$ and $-6\%$
respectively.\cite{Czarnecki:1996fw}  
$\gamma Z$
mixing along with the anapole moment in Fig.~\ref{fig:ew} is the
largest effect.  It effectively replaces the tree level 
$1-4\sw$ in $\alr$ by\cite{Czarnecki:1996fw}  
\ba
1-4\kappa(0) \sw(m_Z)_\msbar
\ea
where 
\ba
\kappa(0) = 1.0301 \pm 0.0025
\ea
represents a 3\% shift in the effective $\sw$ due to loop effects
illustrated in Fig.~\ref{fig:ew}.  That $+3\%$ increase in the
effective $\sw$ appropriate for low $Q^2$ gives rise to a $-38\%$
reduction in $\alr$.  Interestingly, that reduction actually 
makes E158 more
sensitive to $\sw(m_Z)_\msbar$ as well as ``new physics.''

In the case of very large $Q^2$, appropriate for $e^-e^-$ colliders,
the electroweak radiative corrections will change and must be
reevaluated.  In particular, the $WW$ box diagram gives a large
negative contribution to $\alr$.  The
effects of $\gamma Z$ mixing and anapole moment
can also be very large, but they are  easy to
obtain from the loops in Fig.~\ref{fig:ew}.  One finds for arbitrary
$Q^2$ that they replace $1-4\sw$ in the tree level asymmetry by
\ba
1-4\kappa(Q^2) \sw(m_Z)_\msbar &\equiv& 1-4\sw (Q^2),
\nonumber \\
\kappa(Q^2) = \kappa_f(Q^2) + \kappa_b(Q^2),
\ea
where the subscripts $f$ and $b$ denote fermion and boson loops, and
$\sw(Q^2)$ is a running effective parameter.  In
perturbation theory (i.e. without QCD dressing) 
\ba
 \kappa_f(Q^2) &=& 1-{\alpha\over  2\pi\sw}
\left\{  
{1\over 3} \sum_f \left( T_{3f} Q_f - 2\sw
Q_f^2\right) \right.
\nonumber \\ &&\left. \qquad\times
\left[
\ln {m_f^2\over m_Z^2} 
-{5\over 3} 
+4z_f
+(1-2z_f)p_f\ln {p_f+1\over p_f-1}
\right]
\right\},
\nonumber \\
z_f &\equiv& {m_f^2\over Q^2}, \qquad p_f \equiv \sqrt{1+4z_f},
\ea
with $T_{3f}=\pm 1/2$, $Q_f=$ fermion charge, and the sum is over all
fermions; 
\ba
\kappa_b(Q^2) &=&  1-{\alpha\over 2\pi\sw}\left\{
 - {42\cw+1\over 12}\ln \cw
 + {1\over 18}\right.
\nonumber \\
&&
 -\left({p\over 2}\ln{p+1\over p-1}-1\right)
\left[
(7-4z)\cw +{1\over 6}(1+4z)
\right]
\nonumber \\
&&\left.
 -z \left[
{3\over 4} - z 
+\left(z-{3\over 2}\right)p \ln {p+1\over p-1} 
+z \left( 2-z\right) \ln^2 {p+1\over p-1}
\right]\right\},
\nonumber \\
z &\equiv & {m_W^2\over Q^2}, \qquad p \equiv  \sqrt{1+4z}.
\ea
(Note, Eqs.~(28) and (29) of ref.~\citenum{Czarnecki:1998xc} 
contain misprints in the
$\kappa(Q^2)$ expressions.)

\begin{figure}[htb]
\hspace*{-0mm}
\psfig{figure=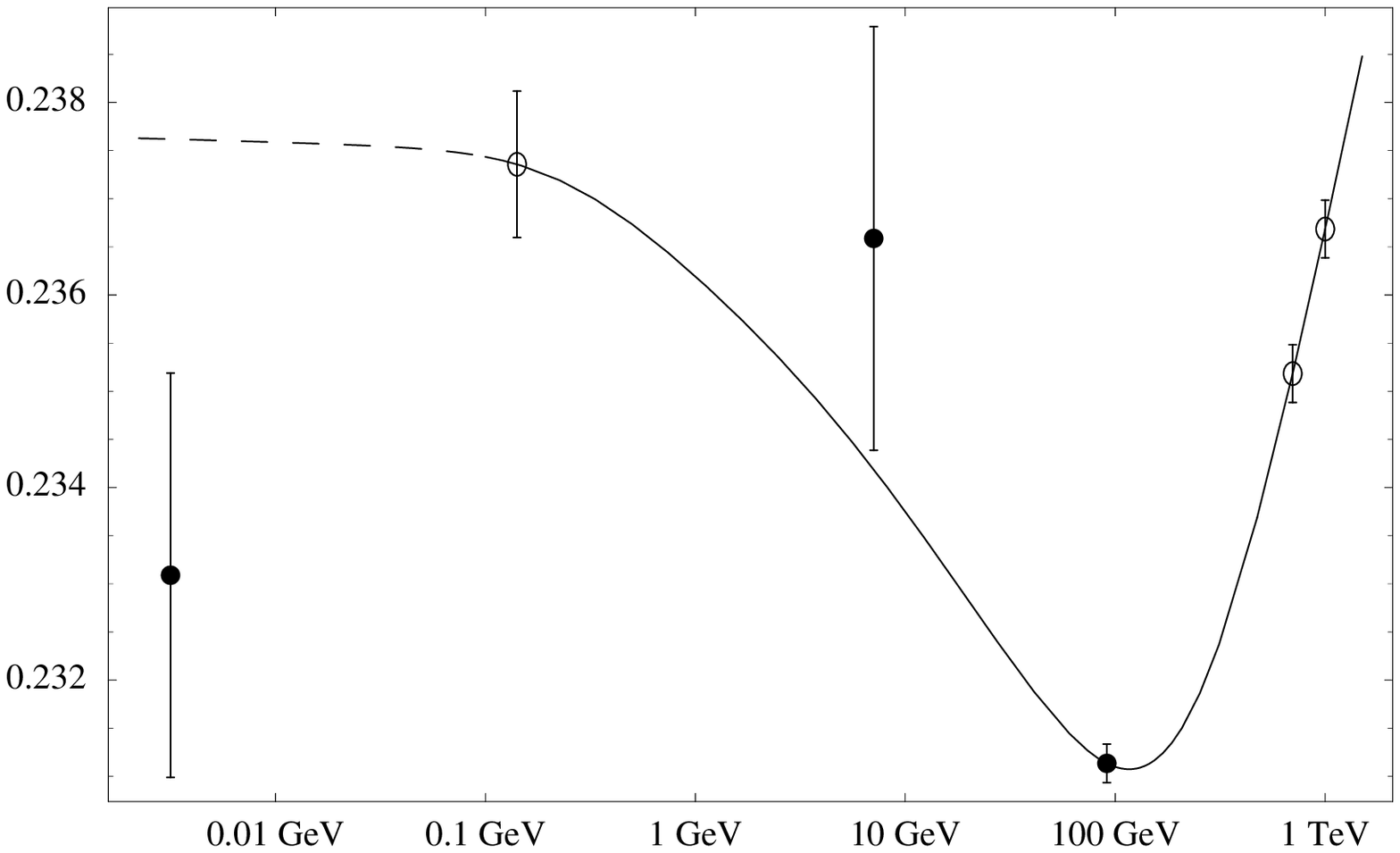,width=120mm,bbllx=72pt,bblly=220pt,%
      bburx=540pt,bbury=580pt} 
\setlength{\unitlength}{1mm}

\begin{picture}(0,0)
\put (18,34) {APV}
\put (48,75) {E158}
\put (80,63) {$\nu N$}
\put (93,23) {$Z$ pole}
\put (102,53) {$e^-e^-$}
\put (105,65) {$e^+e^-$}
\put (124,12) {$Q$}
\put (1,86) {$\sw(Q^2)$}
\end{picture}
\vspace*{-2mm}
\noindent 
\fcaption{Predicted running of $\sw(Q^2)$ and evidence from existing 
experiments (dark circles) along with expectations from 
potential future M{\o}ller and $e^+e^-$
asymmetry measurements at $\sqrt{s}=1$ TeV.}

\label{fig:asym}
\end{figure}

In Fig.~\ref{fig:asym} we illustrate the expected dependence of $\sw(Q^2)$
on $Q$ and show how well it has already been measured for several
$Q^2$.  We also illustrate the approximate 
potential of E158 and future $e^-e^-$
and $e^+e^-$ collider measurements at $\sqrt{s} = 1$ TeV.  
One notices a $2\sigma$ discrepancy in the atomic parity violation
result as compared with Standard Model expectations.  That issue could
be resolved or made even more interesting by results from E158 at SLAC.

\begin{figure}[htb]
\hspace*{-0mm}
\psfig{figure=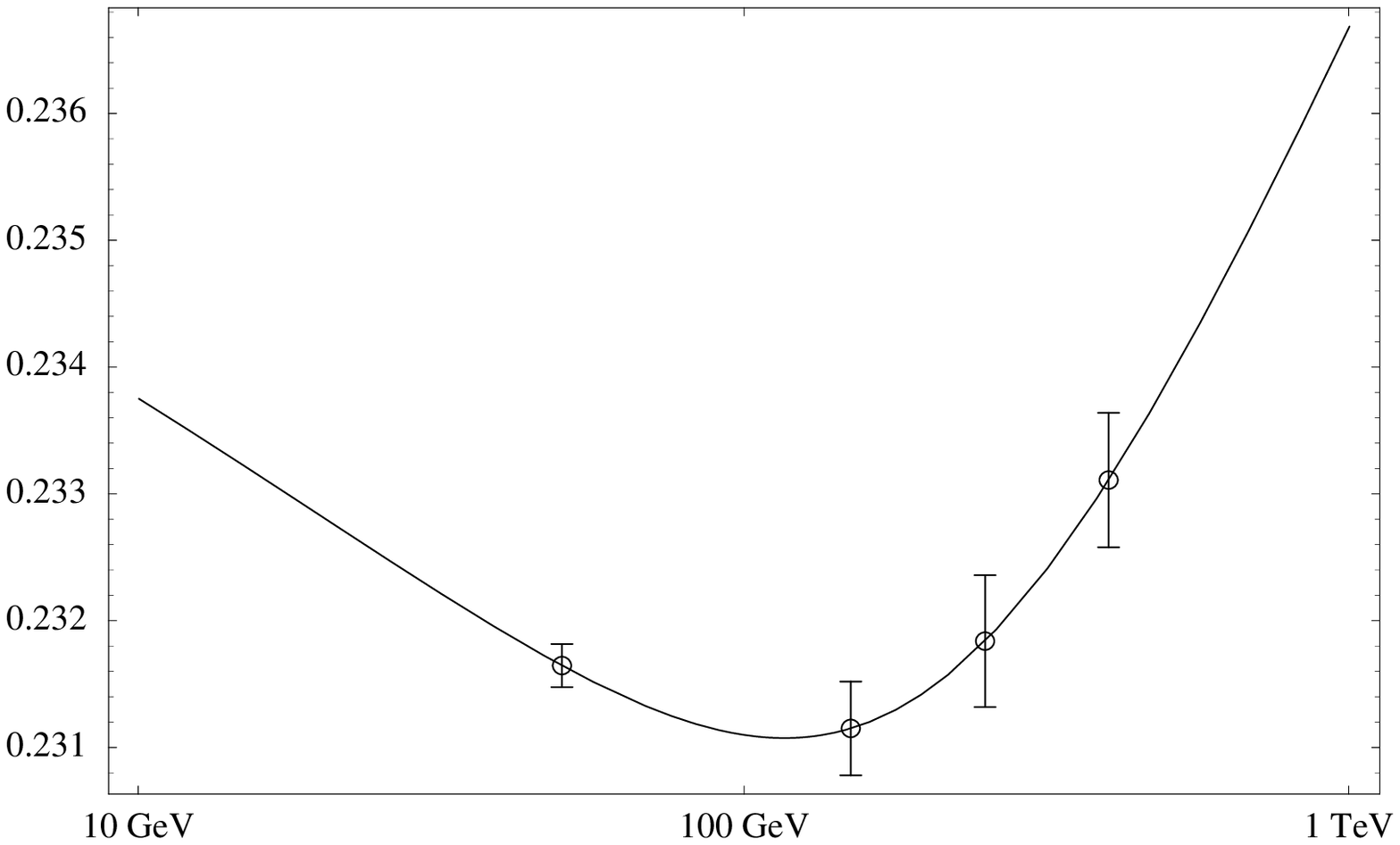,width=120mm,bbllx=72pt,bblly=220pt,%
      bburx=540pt,bbury=580pt} 

\setlength{\unitlength}{1mm}

\begin{picture}(0,0)
\put (124,12) {$Q$}
\put (1,86) {$\sw(Q^2)$}
\end{picture}

\vspace*{-9mm}

\noindent 
\fcaption{Running of $\sw(Q^2)$ compared with potential 
future $e^-e^-$ collider measurements at $\sqrt{s} = 1$ TeV.} 
\label{fig:runee}
\end{figure}

In the case of $e^-e^-$ collider studies, one can actually map out the
variation in $\sw(Q^2)$ in a single experiment 
through  measurements at different $\theta$.
We illustrate in Fig.~\ref{fig:runee} the type of running that one is
predicted to find at a $\sqrt{s} = 1$~TeV $e^-e^-$ collider.  Notice,
that by 
going to small angles (low $Q^2$), one can obtain very high precision.
Of course, within the Standard Model, the measurements at different
$Q^2$ would be radiatively 
corrected to provide a single precise determination of
$\sw(m_Z)_\msbar$.  However, demonstrating the running of $\sw(Q^2)$
over a large range in $Q^2$ in a single experiment 
will be an added bonus.

\section{``New Physics'' Effects}
The real utility of high precision $\alr$ measurements away from the
$Z$ pole is to search for or constrain ``new physics.''  A
disagreement with the extracted $\sw(m_Z)_\msbar $ value from $Z$ pole
determinations could signal the presence of additional tree or loop
level neutral current effects.  Examples that have been considered
include $Z'$ bosons, compositeness, anomalous anapole moment effects,
doubly charged scalars $\Delta^{--}$, extra dimensions, 
etc.  For example, if E158 meets
its phase one goal of a $\pm 0.0007$ determination of
$\sw(m_Z)_\msbar$, it will probe the $m_{Z_\chi}$ of SO(10) at about
the 800 GeV level, compositeness at the 10--15 TeV scale, the anapole
moment at $10^{-17}$ cm (or the $X$ parameter\cite{Maks94} at $\pm
0.15$), and $g^2/m^2_{\Delta^{--}}\sim 0.01G_\mu$.

At an $e^-e^-$ collider, the larger value of $s$ would significantly
improve the ``new physics'' reach.  Roughly, at $\sqrt{s} \simeq 500 $
GeV one could do a factor of 10 better in $m_{Z_\chi}$ and
$\Lambda_{\rm comp}$ than E158.  In the case of the doubly charged
Higgs, $g^2/m^2_{\Delta^{--}} \sim 5\times 10^{-5} G_\mu$ would be
probed.  Of course, the sensitivity would further improve as higher
$\sqrt{s}$ values are reached.

Parity violating left-right asymmetries have played key roles in
establishing the validity of the Standard Model.  From the classic
SLAC polarized $eD$ measurement to the $Z$ pole asymmetry, polarized
electron beams have proved their worth. They will continue to provide
valuable tools during the NLC era both in the $e^+e^-$ and $e^-e^-$
modes.  In the case of precision studies of parity violating
left-right scattering asymmetries, short--distance physics up to
${\cal O}$(150~TeV) will be indirectly explored. Even more exciting is
the possible direct detection of new phenomena such as supersymmetry
at these high energy facilities.  If ``new physics'' is uncovered,
polarization will help sort out its properties and decipher its place
in nature.

\nonumsection{Acknowledgments}
This work was supported  by the DOE under grant number
DE-AC02-76CH00016.

\nonumsection{References} 

\end{document}